# Interferences of electrostatic moiré potentials and bichromatic superlattices of electrons and excitons in transition metal dichalcogenides

Qingjun Tong[1,2], Mingxing Chen[3], Feiping Xiao[1], Hongyi Yu[4], and Wang Yao[*,2,5]

[1]School of Physics and Electronics, Hunan University, Changsha 410082, China

[2]Department of Physics, The University of Hong Kong, Hong Kong, China

[3]School of Physics and Electronics, Hunan Normal University, Key Laboratory for Matter Microstructure and Function of Hunan Province, Key Laboratory of Low-Dimensional Quantum Structures and Quantum Control of Ministry of Education, Changsha 410081, China

[4]Guangdong Provincial Key Laboratory of Quantum Metrology and Sensing & School of Physics and Astronomy, Sun Yat-Sen University (Zhuhai Campus), Zhuhai 519082, China

[5]HKU-UCAS Joint Institute of Theoretical and Computational Physics at Hong Kong, China

* Email: wangyao@hku.hk

**Abstract:**

Recent experimental progresses have demonstrated the great potential of electronic and excitonic moiré superlattices in transition metal dichalcogenides (TMDs) for quantum many-body simulations and quantum optics applications. Here we reveal that the moiré potential landscapes in the TMDs heterostructures have an electrostatic origin from the spontaneous charge transfer across the heterointerfaces dependent on the atomic registry. This allows engineering tunable multi-chromatic superlattices through the interference of moiré potentials from independently configurable heterointerfaces in multilayers. We show examples of bichromatic moiré potentials for valley electrons, holes, and interlayer trions in $MX_2/M'X'_2/MX_2$ trilayers, which can be strain switched from multi-orbital periodic superlattices to quasi-periodic disordered landscape. The trilayer moiré also hosts two independently configurable triangular superlattices of neutral excitons with opposite electric dipoles. These findings greatly enrich the versatility and controllability of TMDs moiré as a quantum simulation platform.

## I. Introduction

The beating of misaligned lattice periodic structures of adjacent 2D crystals leads to the formation of moiré pattern, which has emerged as an exciting venue to explore fundamental physics and to tailor material properties in van der Waals structures. A plethora of novel electronic phenomena are observed in graphene moiré superlattices including the fractal quantum Hall effect [1-4], and the emergent correlation phenomena in flat mini-bands at the magic angle twisting [5-13]. Moiré formed in heterostructures of transition metal dichalcogenides (TMDs) proves to be another highly viable platform for exploration of the exciting superlattice physics. Recent experiments have discovered remarkably rich correlated insulating phases when the moiré superlattices are loaded with electrons at the various filling factors [14-21], showing a great potential for quantum simulation of Hubbard model physics. TMDs heterostructures also host the interlayer valley exciton, a long-lived optical excitation that features strong many-body interaction through its electrical dipole. These excitons also experience sizable superlattice potentials in the moiré [22-27], which can be exploited as quantum emitter arrays with protected polarization selection rules [27-29], as well as for quantum simulation of many-body physics with the unique optical addressability.

The atomic structure of TMDs determines that moiré between two heterolayers realizes a simple triangular superlattice of one trapping site per supercell in general [14-21]. With the robust many-body correlations demonstrated in the simple TMDs moiré, more versatile superlattice landscapes can be desirable for exploring a broad range of phenomena based on the moiré systems. In the use of cold atom systems for quantum simulation of condensed matter phenomena, the interference of optical lattices of different wavelength has been exploited to realize tunable complex landscapes, such as the quasi-periodic bichromatic lattices which have provided an important arena to study Anderson localization [30] and many-body localization [31].

Here we show this strategy can be borrowed to engineer versatile superlattice landscapes in TMDs heterostructures. We find that moiré potentials for the K valley electrons, holes and excitons in TMDs heterostructures are predominantly from the spontaneous electrical polarization (interlayer charge transfer) across the heterointerface that are spatially modulated by the textures of atomic registries. Through the charge and electric dipole, the valley carriers and excitons experience non-locally the electrostatic moiré potentials produced at all adjacent heterointerfaces. The beating of the moiré potentials from two separately configurable heterointerfaces in a trilayer thus lead to tunable bichromatic superlattices. These findings, drawn from first-principles calculations of charge distributions, electrostatic potentials and band structures in heterobilayers, are corroborated by direct computation of band edge energies in trilayers with two heterointerfaces of various R-stacking registries. We show examples of bichromatic moiré potentials for electrons, holes, and interlayer trions in $MX_2$/$M'X'_2$/$MX_2$ trilayers, which can be switched from multi-orbital periodic superlattices (i.e. multiple trapping sites per supercell) to quasi-periodic disordered landscape by straining the bottom layer. The trilayer moiré also hosts two independently configurable triangular superlattices of neutral excitons with opposite electric dipoles. By adding more misaligned layers, multi-chromatic landscapes can be created.

## II. Characteristics of concatenated moiré pattern

We first describe the spatial textures of atomic registries in the trilayer. Take the middle layer as the reference, with which the top/bottom layer has small lattice mismatch $\delta^{t/b}$ and small twisting angle $\theta^{t/b}$. These lead to two pair-wise long-wavelength moiré patterns: moiré-t between top and middle layers, and moiré-b between middle and bottom layers. Each of these moiré patterns is characterized by the mapping function between lateral position $\mathbf{R}$ and the local stacking registry $\mathbf{r}^{t/b}$ (i.e. the in-plane displacement of a metal atom in top/bottom layer from a nearest one in middle layer) [32]. The trilayer atomic texture is then described by two pair-wise local stacking registries as functions of location: $\mathbf{r}^t(\mathbf{R})$ accounting for the moiré-t, and $\mathbf{r}^b(\mathbf{R})$ accounting for the moiré-b, which uniquely specify how the three layers are vertically stacked at any given $\mathbf{R}$.

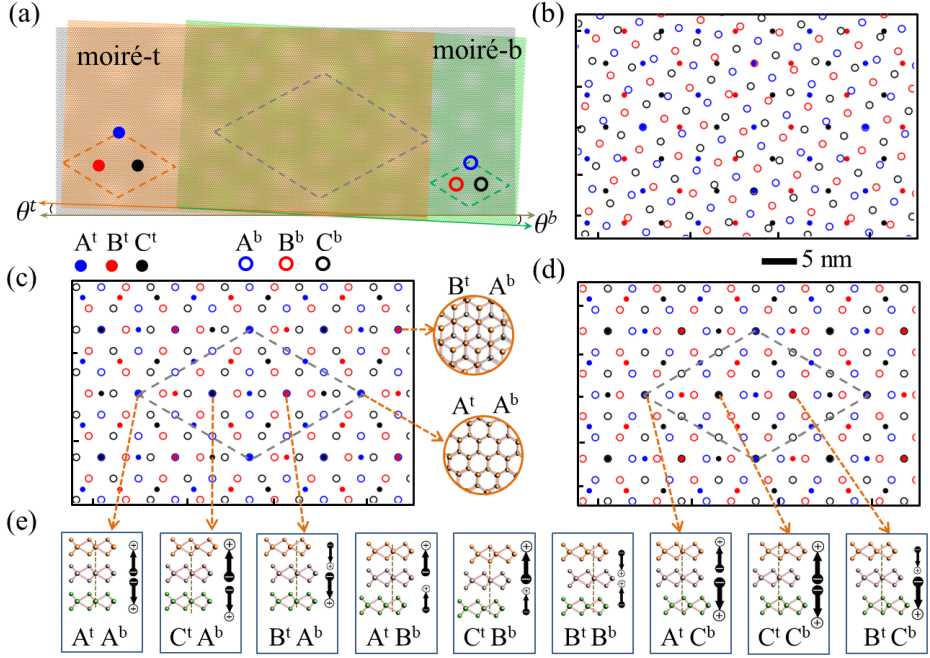

**Figure 1.** (a) A concatenated moiré pattern formed by twisting the top (orange lattice) and bottom (green lattice) layers with respect to the middle (gray lattice) layer by angles $\theta^t = 2°$ and $\theta^b = 3°$. The orange, green, and gray diamonds are supercells of moiré-t between top and middle, moiré-b between middle and bottom layers, and the concatenated moiré from the beating of moiré-t and moiré-b. (b-d) The concatenated moiré visualized by tracing the high-symmetry stacking locals of moiré-t (moiré-b), $A^t$, $B^t$ and $C^t$ ($A^b$, $B^b$ and $C^b$), denoted respectively by blue, red, and black solid (empty) dots. Each of these dots encloses tens to hundreds of atoms. The two commensurate concatenated moirés shown in (c) and (d) are connected by a translation of the bottom layer, while a 2% biaxial strain on the bottom layer leads to the incommensurate one shown in (b). (e) Nine high-symmetry stackings of a trilayer.

Moiré-t and moiré-b each has three high-symmetry stacking locals in its supercell denoted as A, B, C, where the three-fold rotational symmetry of lattice is preserved. $B^{t/b}$ stands for the $M^{t/b}$-$X^m$ stacking, where metal (M) atom of top/bottom layer is vertically aligned with chalcogen (X) atom of middle layer, corresponding to $\mathbf{r}^{t/b} = (\mathbf{a}_1 + \mathbf{a}_2)/3$, with $\mathbf{a}_{1,2}$ being the unit lattice vectors of middle layer. Likewise, $A^{t/b}$ stands for the $M^{t/b}$-$M^m$ stacking ($\mathbf{r}^{t/b} = 0$), and $C^{t/b}$ for the $X^{t/b}$-$M^m$ stacking ($\mathbf{r}^{t/b} = 2(\mathbf{a}_1 + \mathbf{a}_2)/3$). The high-symmetry stacking locals can be used to track the complex atomic registry textures in the trilayers (c.f. Figure 1).

Figure 1a shows an example of trilayer with $\theta^t = 2°$, $\theta^b = 3°$, $\delta^t = \delta^b = 0$, where moiré-t and moiré-b are commensurate with 2:3 ratio of their periods. The A, B, C stacking locals of moiré-t (moiré-b) are denoted by the blue, red, and black solid (empty) dots. The beating of the periodicity of solid dots with that of the empty dots gives the larger scale trilayer textures, which is another moiré pattern by viewing the dots as "artificial atoms". Trilayer therefore features a concatenated moiré structure (Figure 1c). We can identify 9 high-symmetry locals in the trilayer, where three-fold rotational symmetry are preserved (Figure 1e). The trilayer moiré pattern possesses remarkable mechanical controllability, by translating or straining one of its layers (Figures 1b-d).

In long-wavelength moiré pattern, there exists an intermediate length scale ($l$), large compared to lattice constant ($a$) and small compared to the moiré wavelength ($b$). In any region of size $l$, the atomic registry closely resembles a lattice-matched stacking, so the local electronic structures in moiré can be well approximated by band structures of the latter [32,33]. This makes possible quantification of moiré energy landscapes from first-principles band structures of lattice-matched stacking of various registries. The latter gives the variation of band edge energies as functions of stacking registry: $V(\mathbf{r})$. Combined with the mapping between $\mathbf{r}$ and location $\mathbf{R}$ in moiré, the energy landscape for the band edge carriers is given from $V(\mathbf{R}) \equiv V(\mathbf{r}(\mathbf{R}))$.

This local approximation has been the standard approach to determine the moiré potential in long-wavelength bilayer moiré [23,32,33], and can be straightforwardly generalized to trilayer. We perform first-principles calculations for lattice-matched trilayers to find out variation of band edge energies $V(\mathbf{r}^t, \mathbf{r}^b)$ as functions of the two pair-wise stacking registries $\mathbf{r}^t$ and $\mathbf{r}^b$. The moiré potential in trilayer is then given from $V(\mathbf{R}) \equiv V(\mathbf{r}^t(\mathbf{R}), \mathbf{r}^b(\mathbf{R}))$.

**III. Electrostatic origin of moiré potentials**

Moiré potentials in the van der Waals structures can have various origins. The interlayer single particle hopping of the band edge electrons at K valleys has a registry dependent amplitude [33-35], and the resultant spatially varying layer hybridization is underlying the formation of superlattice minibands in twisted graphene [5-13,34]. Moiré patterns with multiple misaligned interfaces have also been exploited to tailor complex electronic structures in graphene [36-39]. In TMDs heterobilayers, however, the interlayer hopping is small for the K valleys where the band edge wavefunctions are primarily localized on the metal layer. In the presence of a much larger band-offset, the major effect of this single particle hopping is a second order energy shift of the band edges.

It has also been noted that in R-stacking TMDs homobilayer, out-of-plane electric polarization can spontaneously emerge, where the spatial symmetry dictates that the polarization has a spatially varying sign in the moiré [40,41]. This can also lead to a superlattice potential on the electrons and the dipolar excitons through the Coulomb interaction. Such an interfacial electrostatic potential has been discussed in van der Waals heterostructures involving hexagonal boron nitride [42,43]. Besides, in moiré with very large wavelength, lattice reconstruction becomes significant, forming stacking domains and dislocation networks where resultant strain pattern can modulate the local band energies [41,44]. Here we do not consider this regime of lattice reconstruction.

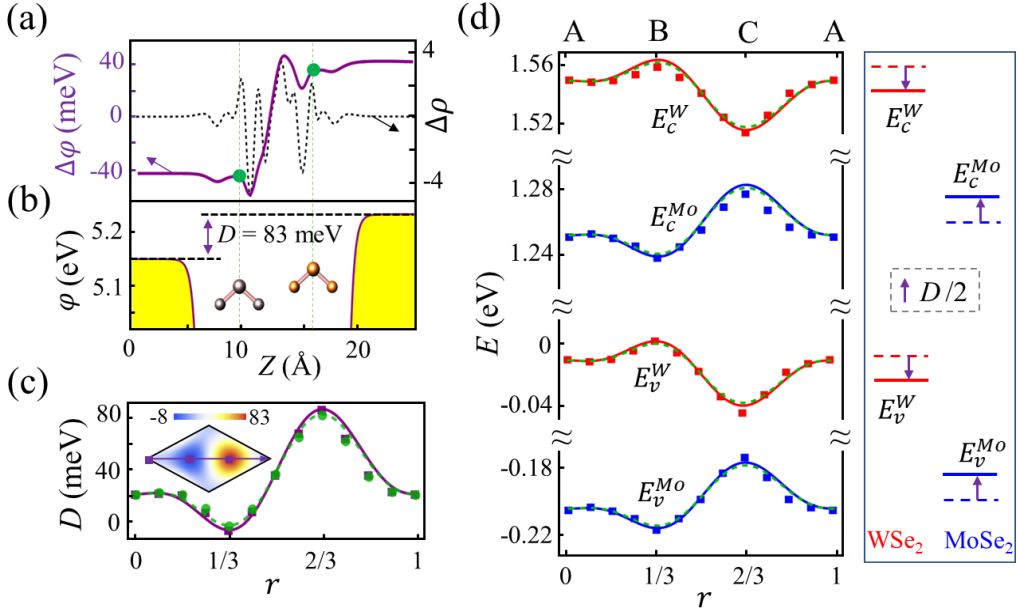

**Figure 2.** (a) Plane-averaged charge density difference $\Delta\rho \equiv \rho_{BL} - \rho_W - \rho_{Mo}$ (in unit of $10^{-4} e/\text{Å}^3$). $\rho_{BL}$ is the charge density of the WSe$_2$/MoSe$_2$ heterobilayer for C configuration, and $\rho_{W/Mo}$ is that of the pristine monolayer. $\Delta\varphi$ is the electrostatic potential generated by $\Delta\rho$, obtained by solving the Poisson equation. $\Delta\varphi$ at the Mo and W atom planes (green dots) are already approaching the values at vacuum levels on the two sides. (b) The electrostatic potential $\varphi$ directly obtained from the first-principles calculations including the contributions from the ionic and Hartree potentials. The charge redistribution $\Delta\rho$ results in a built-in voltage $D$ between the vacuum levels on the two sides. $D$ can also be extracted from the values of $\Delta\varphi$ at the vacuum regions, and from (a) one finds $D \cong \Delta\varphi(\text{Mo}) - \Delta\varphi(\text{W})$. (c) $D$ as a function of interlayer registry $\mathbf{r}$. The purple squares are the vacuum level difference, and the green dots are the difference of $\Delta\varphi$ taken at the two metal layers. The lines are the fitting using Eq. (1) with $\{D_0, D_{+1}, D_{-1}\} = \{20, -8, 83\}$ and $\{20, -4, 79\}$ meV respectively. Inset shows $D(\mathbf{r})$ in the whole unit cell, and the horizontal axis in (c) and (d) corresponds to its long diagonal. (d) The squares plot the type-II band edge energies at K point as functions of interlayer registry $\mathbf{r}$ from first-principles band structures. The solid and dashed lines are $\pm D(\mathbf{r})/2$ plotted using Eq. (1) with the same fitting parameters in (c).

To elucidate the origin of moiré potential for the K valley carriers in the TMDs heterobilayers, we perform first-principles calculations to calibrate the conduction band minima (CBM), valence band maxima (VBM), and the charge distribution under the various stacking registries (see Methods section). In Figure 2a, the dotted curve shows the plane-averaged charge density distribution $\Delta\rho$ due to the interlayer coupling in a MoSe$_2$/WSe$_2$ bilayer of C configuration. The electrostatic potential $\Delta\varphi$ as a function of the vertical coordinate z is then computed from $\Delta\rho$ using the Poisson equation (solid curve in Figure 2a). $\Delta\varphi$ has a sizable drop across the van der Waals gap between the two layers, and the saturated values in the vacuum have a difference of $D \sim 83$ meV on the two sides of the bilayer. Notably, the values of $\Delta\varphi$ at the two metal planes already approach the vacuum values of $D/2$ and $-D/2$ respectively (c.f. green dots on the curve of $\Delta\varphi$ in Figure 2a). $D$ is referred below as the built-in voltage, as in the context of a PN junction.

Figure 2c plots the built-in voltage $D$ as a function of stacking registry $\mathbf{r}$, which varies significantly in a range from -8 to 83 meV. Dictated by the three-fold rotational symmetry and the lattice periodicity, we find $D(\mathbf{r})$ can be well described in terms of the lowest several harmonics [23,45],

$$D(\mathbf{r}) = D_0 f_0(\mathbf{r}) + D_{+1} f_{+1}(\mathbf{r}) + D_{-1} f_{-1}(\mathbf{r})] \qquad (1)$$

where $f_m(\mathbf{r}) = \frac{1}{9}\left|e^{-i\mathbf{K}\cdot\mathbf{r}} + e^{-i\left(\hat{C}_3 \mathbf{K}\cdot\mathbf{r} - m\frac{2\pi}{3}\right)} + e^{-i\left(\hat{C}_3^2 \mathbf{K}\cdot\mathbf{r} + m\frac{2\pi}{3}\right)}\right|^2$ with $\mathbf{K}$ being the wavevector at the corner of the monolayer Brillouin zone. The three parameters $\{D_0, D_{+1}, D_{-1}\}$ can be obtained from the values of $D$ under the high-symmetry A, B and C stackings (c.f. caption of Figure 2). The curves in Figure 2c are the plot of Eq. (1) using these parameters, which are in excellent agreement with the first-principles results shown by the squares and dots.

The stacking dependence of charge redistribution arises from the different alignment of the metal atoms and chalcogen atoms from the two layers. With a larger atomic shell number, the W atom is easier to lose electrons to the Se atom than Mo atom. As a result, for C configuration, in which the W atom is vertically aligned with Se atom of $MoSe_2$, the charge transfer (and $D$) is largest. B configuration has the Mo atom vertically aligned with Se atom of $WSe_2$, and their charge transfer is compensated by the opposite interlayer transfer between the misaligned W atom and Se atom of $MoSe_2$ layer, so the net effect is a small negative $D$. In A configuration, although the interlayer distances from Mo and W to the Se atoms are the same, their different abilities to lose electrons result in an intermediate $D$ value. The interlayer charge redistribution is also influenced by the interlayer distance, which is stacking registry dependent as well (see supplementary Figure 7). In our first-principles calculations, we have allowed structural relaxation to find the interlayer distance at each stacking registry $r$, and the effect of the registry dependence of interlayer distance on $D(r)$ has been accounted.

For the CBM and VBM at the K valleys, the interlayer hopping is much weaker compared to the band offset, so the Bloch functions are predominantly localized on the metal plane of an individual layer. One thus expects the interlayer charge redistribution will contribute a shift of $D(\mathbf{r})/2$ to the CBM and VBM in the MoSe$_2$, and $-D(\mathbf{r})/2$ in the WSe$_2$ layer. We find this electrostatic contribution predominantly accounts for the band energy variation $V(\mathbf{r})$ as functions of stacking registry. In Figure 2d, the $\pm D(\mathbf{r})/2$ curves plotted using Eq. (1) and parameters fitted in Figure 2c are compared with the CBM and VBM directly extracted from the first-principles band structures, which show excellent agreement. The above conclusions also apply for other heterobilayer combinations, including MoS$_2$/WS$_2$, MoS$_2$/WSe$_2$, and WS$_2$/WSe$_2$, which are presented in supplementary Figures 1-3.

With the knowledge of interlayer charge transfer in heterobilayers, we now examine the MX$_2$/M'X'$_2$/MX$_2$ trilayers. Because of the small magnitude of $\Delta\rho$, the charge transfer at the different heterointerfaces turns out to be independent of each other. The built-in voltage across a trilayer of the pair-wise stacking registries $\mathbf{r}^t$ and $\mathbf{r}^b$ is then given by $D^t(\mathbf{r}^t) - D^b(\mathbf{r}^b)$, where $D^t(D^b)$ is the voltage drop between upper (lower) and middle layer whose dependence on $\mathbf{r}^t$ ($\mathbf{r}^b$) is described by Eq. (1) established from bilayer calculations. Figure 3(a) gives a plot of $D^t(\mathbf{r}^t) - D^b(\mathbf{r}^b)$, with the set of fitting parameters extracted from MoSe$_2$/WSe$_2$ heterobilayer calculations (Figure 2c). These curves based on the heterobilayer fitting equations and parameters are then compared with the first-principles calculations of the vacuum level difference in MoSe$_2$/WSe$_2$/MoSe$_2$ trilayer (squares in Figures 3c and 3d), which show excellent agreement.

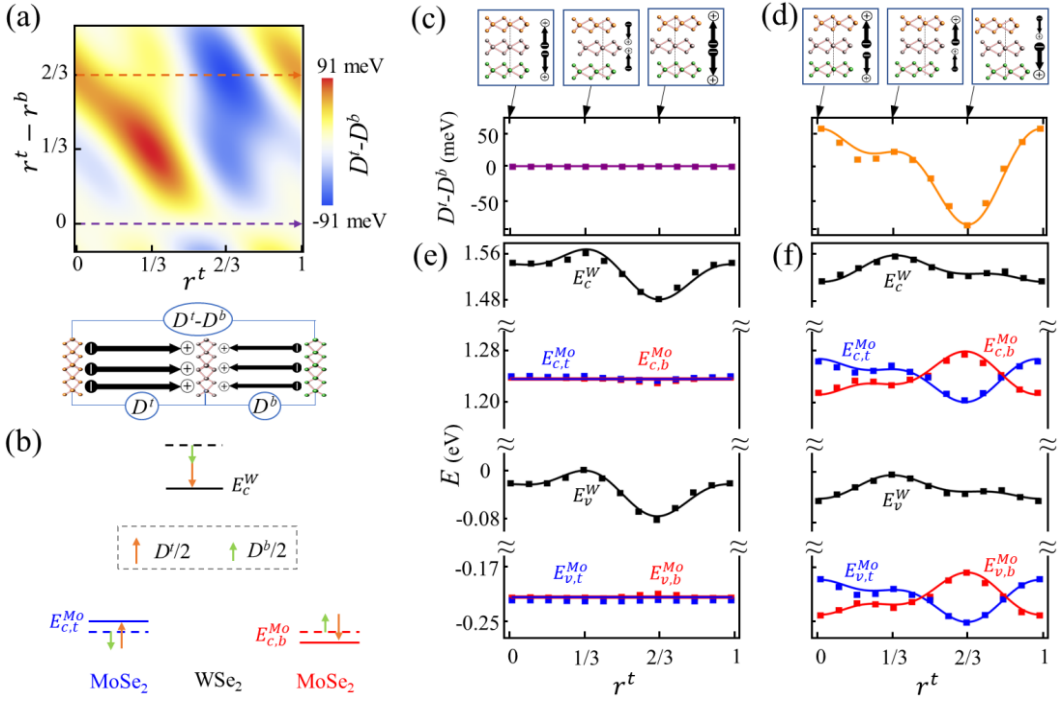

**Figure 3.** (a) $D^t(\mathbf{r}^t) - D^b(\mathbf{r}^b)$ plotted using Eq. (1), which corresponds to the built-in voltage across a MoSe$_2$/WSe$_2$/MoSe$_2$ trilayer, as the lower panel illustrates. (b) Schematics of charge transfer induced shifts of CBM in the three layers. The shifts of VBM have the same pattern. (c,d) The squares give built-in voltage directly extracted from the first-principles calculated vacuum level difference of the electrostatic potential on the two sides of the trilayer, similar to the bilayer case as illustrated in Figure 2b. The curves are $D^t(\mathbf{r}^t) - D^b(\mathbf{r}^b)$ plotted using Eq. (1). (e,f) The squares are the energies of the CBMs and VBMs in the three layers determined from the first-principles band structures. The curves are plotted using Eqs. (2) and (3). The trilayer configurations of various pair-wise interlayer registries ($\mathbf{r}^t$, $\mathbf{r}^b$) are taken on the lower dashed line (for c,e), and the upper dashed line (for d,f) in part (a). All curves are plotted using Eq. (1) with the parameters $\{D_0, D_{+1}, D_{-1}\} = \{20, -8, 83\}$ meV extracted from fitting the heterobilayer data (Figure 2c).

With the staggered band alignment in MoSe$_2$/WSe$_2$/MoSe$_2$ trilayers (Figure 3b), the large band offset between neighboring layers quenches layer hybridization at the CBM and VBM in the K valleys, so the wavefunctions are all localized in individual layers. The charge transfers at the two heterointerfaces lead to an overall energy shift of the CBM and VBM in the MoSe$_2$ layers,

$$V_{c/v}^{Mo,t}(\mathbf{r}^t, \mathbf{r}^b) = -V_{c/v}^{Mo,b}(\mathbf{r}^t, \mathbf{r}^b) = [D^t(\mathbf{r}^t) - D^b(\mathbf{r}^b)]/2, \quad (2)$$

with $t$ and $b$ denoting the top and bottom layer here. In contrast, the energy shift of CBM and VBM in the middle WSe$_2$ layer is given by

$$V_{c/v}^{W}(\mathbf{r}^t, \mathbf{r}^b) = -[D^t(\mathbf{r}^t) + D^b(\mathbf{r}^b)]/2. \quad (3)$$

These are schematically shown in Figure 3b. Like in the bilayer case, we find that this contribution predominantly accounts for the band energy variation in trilayer. In Figures 3e and 3f, the squares are the band edge energies directly extracted from the first-principles band structures for trilayer of various stacking registries $(\mathbf{r}^t, \mathbf{r}^b)$, which agrees remarkably well with Eqs. (2) and (3) adopting the bilayers fitting equations and parameters (curves). In supplementary Figures 4-6, we also present the band edge variations in $MoS_2/WS_2/MoS_2$, $MoS_2/WSe_2/MoS_2$, and $WS_2/WSe_2/WS_2$ trilayers, which can all be accounted by the built-in potentials $D^t$ and $D^b$ from the charge transfers at the corresponding heterointerfaces (c.f. supplementary Figures 1-3).

### IV. Bichromatic moiré potentials for electrons and excitons in the trilayer

In a long-wavelength moiré pattern, the smoothly changing atomic registries $(\mathbf{r}^t, \mathbf{r}^b)$ as a function of position $\mathbf{R}$ then creates a lateral modulation of the local CBM and VBM energies, which becomes the moiré potential. For the electrons and holes in the various bands and layers, the moiré potential manifests as the sum of the contributions from the two heterointerfaces:

$$V(\mathbf{R}) \equiv V\left(\mathbf{r}^t(\mathbf{R}), \mathbf{r}^b(\mathbf{R})\right) = \pm \left[D^t(\mathbf{r}^t(\mathbf{R})) \pm D^b(\mathbf{r}^b(\mathbf{R}))\right]/2$$

The atomic registry textures $\mathbf{r}^t(\mathbf{R})$ and $\mathbf{r}^b(\mathbf{R})$ at the two interfaces have their wavelengths and orientations separately tunable by the twisting angles $\theta^t$ and $\theta^b$, as well as strain. The two contributions $D^t(\mathbf{R}) \equiv D^t(\mathbf{r}^t(\mathbf{R}))$ and $D^b(\mathbf{R}) \equiv D^b(\mathbf{r}^b(\mathbf{R}))$ are thus separately configurable, and their beating can be exploited to engineer versatile energy landscapes for carriers and excitons.

**Table I.** Moiré superlattice potentials for electrons ($V_e^{t/b}$), holes ($V_h$), neutral excitons ($V_{X0}^{t/b}$), and trions ($V_{X-}$) in $MoSe_2/WSe_2/MoSe_2$ trilayers. The superscripts t and b denote electron (electron component of $X_0$) being in the top and bottom $MoSe_2$ layers. $D^t(\mathbf{R})$ and $D^b(\mathbf{R})$ are respectively the electrostatic potential produced by the charge transfer at the first and second heterointerfaces, which are separately configurable. The beating between $D^t(\mathbf{R})$ and $D^b(\mathbf{R})$ can then be exploited to engineer versatile complex superlattices.

| $V_e^t$ | $V_e^b$ | $V_h$ | $V_{X0}^t$ | $V_{X0}^b$ | $V_{X-}$ |
|---|---|---|---|---|---|
| $\dfrac{D^t(\mathbf{R}) - D^b(\mathbf{R})}{2}$ | $\dfrac{D^b(\mathbf{R}) - D^t(\mathbf{R})}{2}$ | $-\dfrac{D^t(\mathbf{R}) + D^b(\mathbf{R})}{2}$ | $D^t(\mathbf{R})$ | $D^b(\mathbf{R})$ | $\dfrac{D^t(\mathbf{R}) + D^b(\mathbf{R})}{2}$ |

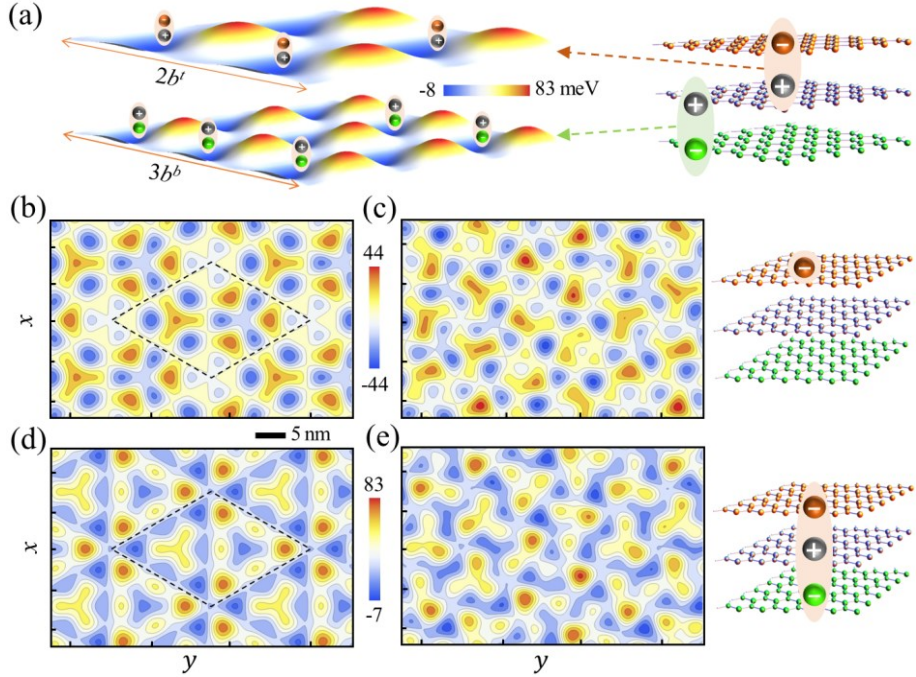

**Figure 4.** (a) Moiré potentials for neutral excitons in a MoSe$_2$/WSe$_2$/MoSe$_2$ trilayer with twisting angles $\theta^t = 2°$ and $\theta^b = 3°$ (c.f. Figure 1a). The two species of interlayer excitons with opposite electric dipoles experience distinct triangular superlattices that can be independently configured through $\theta^t$ and $\theta^b$. (b) Moiré potential for electrons in the top layer realizes a multi-orbital superlattice (i.e. multiple trapping sites per supercell). The potential for electrons in the bottom layer has the same profile with an opposite sign, i.e. $V_e^b(\mathbf{R}) = -V_e^t(\mathbf{R})$. (d) Moiré potential for interlayer trions which realizes another complex superlattice. The potential for holes in the WSe$_2$ layer has the same profile with an opposite sign, i.e. $V_h(\mathbf{R}) = -V_{X-}(\mathbf{R})$ (c.f. Table I). A biaxial tensile strain of 2% on the bottom layer changes the energy landscapes in (b) and (d) to the quasi-periodic disordered ones in (c) and (e) respectively.

In MoSe$_2$/WSe$_2$/MoSe$_2$ with the staggered type-II band alignment (Figure 3b), the low-energy carriers are electrons in the top and bottom MoSe$_2$ layers, and holes in middle WSe$_2$ layer. For neutral excitons, the lowest energy ones are of the interlayer configurations which can reside in the top-middle layers, or the middle-bottom layers, featuring the opposite electric dipoles. Moreover, the neutral excitons can further bind an electron to form a negatively charged exciton (trion) with its three constituents in three separate layers. Table I lists the moiré potentials for these carriers and excitons. For excitons and trions, we do not consider here the spatial variation in binding energy as the Bohr radius is large compared to the small variation of interlayer distance in the moiré [22].

Figure 4a plots the moiré potentials for the neutral excitons in a MoSe$_2$/WSe$_2$/MoSe$_2$ trilayer. The moiré potential for these dipolar interlayer excitons is determined by the charge transfer between the two hosting layers only (see Table I), not affected by the third layer. Therefore the interlayer excitons residing in the top-middle layers and bottom-middle layers experience triangular superlattice potentials of independently configurable orientation and wavelength. The one shown is for twisting angles $\theta^t = 2°$ and $\theta^b = 3°$.

Figure 4b plots the moiré potential for electrons in the top MoSe$_2$ layer in the trilayer moiré of $\theta^t = 2°$ and $\theta^b = 3°$. The beating of the commensurate $D^t(\mathbf{R})$ and $D^b(\mathbf{R})$ creates a multi-orbital periodic superlattices (i.e. multiple trapping sites per supercell). Note that electrons in the bottom layer experience moiré potential having the same profile with an opposite sign, i.e. $V_e^b(\mathbf{R}) = -V_e^t(\mathbf{R})$, where the energy minima and maxima are switched (see Table I). There are 9 (4) local minima for electron in the top (bottom) layer. The moiré potentials for holes and trions have a different landscape from those of electrons, as they are created by the opposite beating pattern between $D^t(\mathbf{R})$ and $D^b(\mathbf{R})$. Figure 4(d) plots the moiré potential for interlayer trion, while that for holes has the same profile with an opposite sign, i.e. $V_h(\mathbf{R}) = -V_{X^-}(\mathbf{R})$ (see Table I). Applying a biaxial strain on the bottom layer transforms the commensurate concatenated moiré into an incommensurate one (c.f. Figure 1b), where the moiré potentials for electrons and trions become the disordered ones as shown in Figures 4c and 4e. Such quasi-periodic moiré potentials are expected at general twisting angles.

These versatile tunable moiré potential landscapes in the heterotrilayer TMDs point to a viable platform to study the transport and many-body phenomena of the interacting electrons and excitons, which also feature the unique optical addressability of their valley and spin degrees of freedom [23]. The complex trilayer moiré can be exploited to simulate the multi-orbital Hubbard physics, where orbital degree of freedom plays key roles in the diverse properties of transition-metal-based materials, such as metal-insulator transition, high-temperature superconductivity, and colossal magnetoresistance [46]. Hetero-strain applied through a substrate can tune the moiré potential from the multi-orbit periodic lattice to the quasi-periodic disordered landscape, which can be exploited for studying Anderson localization and many-body localization [30,31]. The double-layer superlattices of neutral excitons can also be exploited as configurable quantum emitter arrays where the repulsive interaction between the layers helps to lift the degeneracies on demand for onset of the single photon statistics. The remarkable tunability of the landscapes by strain and interlayer bias further imply possibility towards device applications based on the electric and mechanical controls of these phenomena for versatile electronics and optoelectronics based on the van der Waals heterostructures.

**Methods**

First-principles calculations were performed using the Vienna *ab initio* Simulation Package [47]. The exchange correlation functional was approximated by the generalized gradient approximation as parametrized by Perdew, Burke, Ernzerhof [48], and pseudopotentials were constructed by the projector augmented wave method [49]. A uniform Monkhorst-Pack *k*-mesh of $15 \times 15 \times 1$ was adopted for integration over the Brillouin zone. A plane-wave cutoff energy of 500 eV was used for all the calculations. The convergence criterion of electronic self-consistent calculations was set as $10^{-6}$ eV. Van der Waals dispersion forces between the TMDs layers were accounted for through the optB88-vdW functional by using the vdW-DF method [50]. During the structural relaxations, the in-plane coordinates of the atoms were frozen (lattice constant of 3.317 Å), while the out-of-plane components were allowed to relax for each translation configuration. We use a slab to model the heterobilayers as well as heterotrilayers. In order to avoid the long-range dipole interaction between the polar slabs, we then double the unit cell by placing two slabs of the heterostructures mirror symmetrically along *z* direction with the vacuum region of about 15 Å [51]. This modeling allows us to obtain well-converged vacuum levels related to the two sides of the heterobilayers and heterotrilayers (Figure 2b). In the plots of band edge energies (Figures 2d, 3e and 3f), the average value of the two vacuum levels was aligned between the configurations of various stacking registries.

**Acknowledgements:** The work is supported by the Research Grants Council of Hong Kong (17312916), Seed Funding for Strategic Interdisciplinary Research Scheme of HKU, National Natural Science Foundation of China (Grant No. 11904095, 11774084, U19A2090, and 91833302), Project of Educational Commission of Hunan Province of China (Grant No. 18A003) and the Fundamental Research Funds for the Central Universities from China.